# Estimation of Individual Micro Data from Aggregated Open Data

Han-mook Yoo, Han-joon Kim
School of Electrical and Computer Engineering
University of Seoul, Republic of Korea
{yhm7305, khj}@uos.ac.kr

Jonghoon Chun
School of Software Convergence
Myongji University, Republic of Korea
jonghoonchun@gmail.com

*Abstract*— **In this paper, we propose a method of estimating individual micro data from aggregated open data based on semi-supervised learning and conditional probability. Firstly, the proposed method collects aggregated open data and support data, which are related to the individual micro data to be estimated. Then, we perform the locality sensitive hashing (LSH) algorithm to find a subset of the support data that is similar to the aggregated open data and then classify them by using the Ensemble classification model, which is learned by semi-supervised learning. Finally, we use conditional probability to estimate the individual micro data by finding the most suitable record for the probability distribution of the individual micro data among the classification results. To evaluate the performance of the proposed method, we estimated the individual building data where the fire occurred using the aggregated fire open data. According to the experimental results, the micro data estimation performance of the proposed method is 59.41% on average in terms of accuracy.**

## I. Introduction

Due to the advent of the Fourth Industrial Revolution, the importance of analytics to open data is growing globally. However, many of the data provided by current open data portals are aggregated, and thus it is not easy to analyze the open data properly. This is because opening the data can be associated with capacity constraints and personal privacy issues. Most of public institutions have an enormous amount of micro data that can be made available to the public [12], and however they are too large to be released as they are. Furthermore, when processing the open data containing individual or personal information, we must comply with the privacy law [13]. For example, the aggregated fire data does not contain specific location information for the fire site. As a result, only statistical analysis of the frequency of fire is possible, and it is impossible to construct a fire prediction model at a specific building level. In this paper, to overcome the limitations of the aggregated open data, we propose a method of estimating the individual micro data from the aggregated open data based on semi-supervised learning and conditional probability. The proposed method firstly collects the aggregated open data and the support data, which are related to the individual micro data to be estimated. For example, to identify a particular building where the fire actually occurred, we collect the aggregated open data related to the fire and the support data with building information covering the relevant area. Then, we perform the locality sensitive hashing (LSH) algorithm to find a subset of the support data that is similar to the aggregated open data and then classify them by using the Ensemble model, which is generated by semi-supervised learning. Finally, we use conditional probability to estimate the individual micro data among the classification results by discovering the most suitable record for the probability distribution of the individual micro data. In this way, the proposed method is able to estimate individual microdata from aggregated fire data, and thus it can significantly enhance the utility of the open data.

## II. Background

### A. Locality Sensitive Hashing

The locality sensitive hashing (LSH) is a technique that uses hash functions to select the data whose similarity is greater than a user-defined threshold [1]. This is done by dividing the features of the data into bands and hashing the rows belonging to the bands. Figure 1 shows the process of performing the LSH. If the hash values of rows belonging to the second band are in the same bucket, then LSH says that $Data_1$ and $Data_2$ are similar. The performance of the LSH is determined by the number of bands and rows.

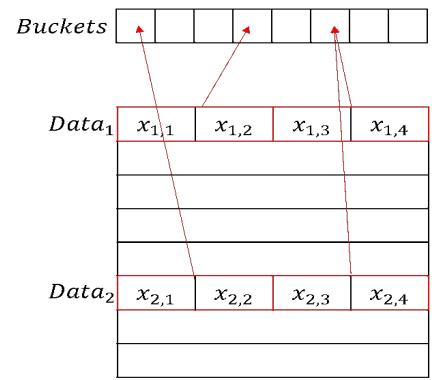

Fig. 1. LSH process with 2 bands and 2 rows

In Equation (1), *S* represents the similarity between two data, and *r* and *b* are the number of rows and bands, respectively. The equation implies that higher *r* and lower *b* values lower the probability that two similar data share the same bucket.

$$\text{Probability of sharing buckets} = 1 - (1 - S^r)^b \quad (1)$$

LSH differs in the way data is defined, depending on the method used to define the similarity between the data [2]. In this paper, we use Jaccard similarity based on non-Euclidean distance.

*B. Expectation Maximization algorithm*

The Expectation-Maximization (EM) algorithm is an algorithm that decomposes a given set of data using the maximum likelihood and maximum posterior probability in a probability model that contains latent variables [3]. The EM algorithm performs repeatedly clustering with two steps: E-step and M-step. The E-step calculates the expected value of the latent variable using the parameters of the given data, and the M-step estimates the parameters using the expectation value of the latent variable from the E-step. The procedure of the above EM algorithm is as follows

1. Given a set of data $x = \{x_1, x_2, x_3, \ldots, x_n\}$ and probability model $p(x, z|\theta)$ that considers latent variable $z$.
2. Initialize parameter $\theta$.
3. (E-step) Estimate the latent variable $z^i$ using the given $\theta^{i+1}$ at every ith step.
4. (M-step) Estimate the parameter $\theta^{i+1}$ using the estimated hidden variable.
5. Repeat E-step and M-step until parameters converge.

Fig. 2. Process of EM algorithm

### III. MOTIVATION EXAMPLE

In this section, we describe an example scenario for outlining how to estimate the individual micro data from the aggregated open data. Suppose we want to use the aggregated open data to find the individual building data where the fire occurred. To find the individual micro data, we need data related to the fire and building information; these are the aggregated open data and support data, respectively. Features of the aggregated open data can be date, time, injury, building_structure, and so on. Features of the support data can be location, building_structure, building_space, permission_date, and so on. Firstly, to estimate the original building data where the fire occurred, we find a subset of the support data that is similar to the aggregated fire open data. In order to find similar data, the aggregated open data and support data must have common features, and in that example, the common feature can be the "building structure"

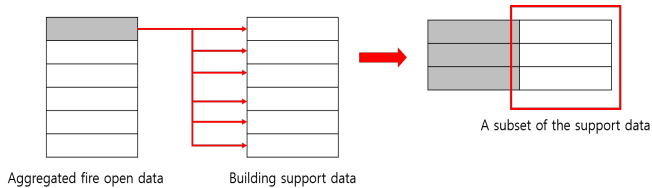

Aggregated fire open data    Building support data    A subset of the support data

Fig. 3. Building a subset of the support data similar to the aggregated fire open data

Figure 3 shows the process of finding a subset of the building support data that is similar to the aggregated fire open data.

According to Figure 3, we compute the similarity between a specific record of aggregated fire open data and the support building data and then compose the most similar building data into a subset. In the result shown in Figure 3, there may be a record of the original building data where the fire occurred (true positive), while some of the subset may be records of the original building data where the fire not occurred (false positive). Therefore, we should find the original building data where the fire occurred among the subset of the support data. For this purpose, we learn a model that classifies two classes (positive class, negative class).

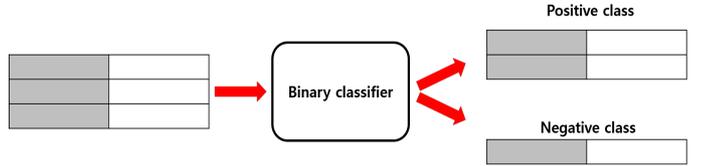

Fig. 4. Classification using binary classifier

Figure 4 shows the classification of the subset of the support data into two classes using the learned classifier. The details on how to learn the classifier are discussed in Section 4.4. In this process, the individual building data where the fire not occurred is filtered based on probability. However, there may remain many building data in the positive class. To estimate the individual micro data, we should choose one of them. We use the conditional probability to select the most probable data among the data classified in the positive class.

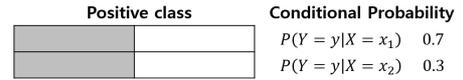

| Positive class | Conditional Probability |
|---|---|
| | $P(Y = y|X = x_1)$   0.7 |
| | $P(Y = y|X = x_2)$   0.3 |

Fig. 5. Data selection using the conditional probability

Figure 5 shows the selection of the most probable data among the data classified in the positive class using the conditional probability. In the conditional probability of this figure, the symbol $y$ is a feature of the aggregated fire open data and the symbol $x_i$ is a feature contained in the support building data. In Figure 5, the conditional probabilities of the two data $x_1$ and $x_2$ are 0.7 and 0.3, respectively. Therefore, we estimate the individual micro data by choosing the most probable 0.7. Like this, it is possible to estimate the individual micro data from aggregated open data.

### IV. ESTIMATION OF INDIVIDUAL MICRO DATA

*A. Overview of estimation of individual micro data*

In this section, we propose a method for estimating the individual micro data from the aggregated open data. Figure 6 shows the overall process of estimating the individual micro data from the aggregated open data. In this figure, the LSH stage performs a similarity calculation between aggregated open data and support data to isolate part of support data; the part of support data means a candidate set of the individual micro data. Due to the nature of open data, the domains of the given data can be different with each other even if they have the common features because they were produced independently. Here, a domain is a set of all the values that a particular feature can have. For this reason, despite the fact that two data are actually similar, there is a possibility that they are classified as false negative by LSH because they have low similarity. Therefore, to compensate for this case, we perform LSH while changing the number of bands and rows differently from the general LSH method which fixes the number of bands and rows.

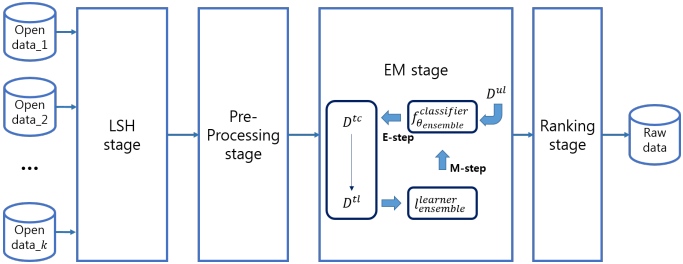

Fig. 6. Overall process of the micro data estimation (restoration)

The second preprocessing stage is a step of preprocessing the data before learning and classifying the result of LSH into the classification model. The third EM stage learns the classification model and uses it to filter the result of LSH. Due to the nature of the purpose of estimating the individual micro data from the aggregated open data, it is costly to obtain many labeled datasets. Therefore, we combine the EM algorithm with the classification model so that the unlabeled data set can be used for classification model learning to perform semi-supervised learning. The final ranking stage assigns ranking by using conditional probability for data predicted as a positive class in classification result.

### B. The LSH stage

As said earlier, to estimate the individual micro data from the aggregated open data, we should find a subset of the support data that is similar to the aggregated open data. For example, estimating the original building data where the fire occurred requires a subset of the support building data which is similar to the building information of the aggregated fire open data. Therefore, we must calculate the similarity for every possible pair between the aggregated open data and the support data. However, there is a problem that the cost of calculation exponentially increases as the number of data increases. To alleviate the problem, we perform the similarity calculation by taking only pairs requiring similarity calculation among all possible pairs between the aggregated open data and the support data using LSH. Here, we should consider that since the open data are generated from the public institutions independently, their domains may be different even if they have the common features. Despite the fact that two data are actually similar, there is a possibility that they are classified as false negative by LSH because they have low similarity. Therefore, we set the similarity threshold of LSH continuously from a high level (0.9) to a low level (0.5). Figure 7 shows the pseudo code for the LSH algorithm performed in this paper.

### C. The preprocessing stage

To estimate the individual micro data from the aggregated open data, the part of the support data generated using the LSH and the individual micro data should form a *one-to-one* relationship with each other. However, the part of the support data and the individual micro data have a *many-to-one* relation.

**Algorithm** LSH
**Input:** Open $data_i$, Support data
**Output:** LSH_candidate_list
1. data1 = Open $data_i$,
2. data2 = Support data
3. band_range= (4, 5, 10, 20)
4. **for** $i$ = 1 to nrow(data1) **do**
5.    band = 4
6.    LSH_candidate_list = LSH(data1[i,], data2, band)
7.    **if** LSH_candidate_list is empty **then**
8.      **for** $j$=2 to length(band_range) **do**
9.        band=band_range[j]
10.        LSH_list=LSH(data1[i], data2, band)
11.        **if** LSH_list is not empty **then**
12.          break
13.        **end if**
14.      **end for**
15.    **end if**
16.    **return** LSH_candidate_list
17. **end for**

Fig. 7. The pseudo code for the LSH algorithm

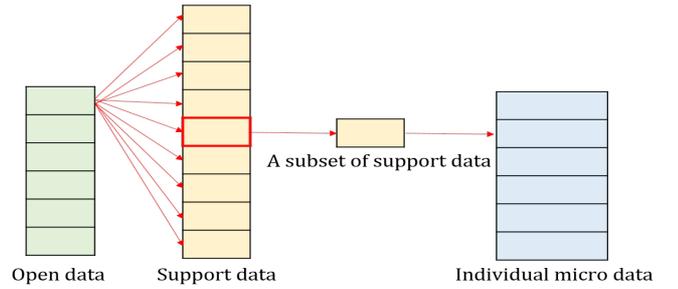

Fig. 8. Ideal individual micro data estimation.

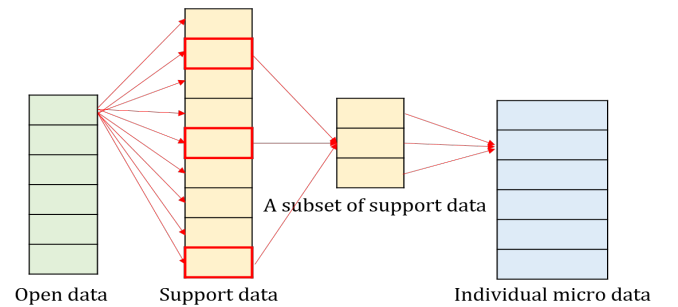

Fig. 9. The actual result of LSH.

Figure 8 shows the ideal individual micro data estimation. In this figure, we calculate the similarity of a record of open data and the support data using LSH. Then, we find the most similar support data and put it into a subset of support data. The generated subset of support data forms a *one-to-one* relationship with the individual micro data. On the other hand, Figure 9 shows the actual result of LSH. In this figure, the subset of support data forms a *many-to-one* relationship with the individual micro data, because there may be several data similar support data in aggregated open data. In this paper, we intend to reduce the subset of support data obtained from the LSH stage by using the classification model, and the preprocessing stage preprocesses the data to learn the classification model.

*D. The EM stage*

In this section, we describe how to reduce the subset of support data by using the classification model learned with the EM algorithm. To perform the supervised learning, we need many labeled datasets for model learning. However, it is practically impossible to obtain many labeled data for the purpose of estimating the individual micro data from the aggregated open data. Thus, we perform the EM-based semi-supervised learning to augment a small number of labeled data with a large number of unlabeled data. The EM algorithm is an algorithm that clusters given data considering latent variables, which can be used as semi-supervised learning in combination with classification models [4]. Normally, the EM algorithm is combined with the Naive Bayes classifier for the semi-supervised learning [4]. The Naive Bayes classifier finds decision boundaries linearly when the features of the input data follow a multinomial distribution [5], which cannot achieve high performance when a given data cannot be linearly classified. Therefore, in this paper, we use the *Ensemble* method combining a number of classification models. The *Ensemble* is a method of combining the multiple weak models to compensate for the limitations of single model prediction, and it has the advantage of typically showing higher performance than when using a single model [6].

The *Ensemble* methods have different types such as bagging, boosting, and stacking [6]. The bagging aims to lower the variance of the classification model, which learns multiple classifiers from the training data and combines the results of each classifier. Here, to combine the generated models, we can use voting, weighted voting, averaging, and so on. Next, the boosting aims to lower the bias of the classification model. Lastly, the stacking aims to lower both the variance and bias of the classification model. In this paper, with considering all three Ensemble methods for the subset of support data, we select the best method and combine it with EM. Figure 10 shows the procedure of combining the Ensemble method with EM and the semi-supervised learning.

1. Learn each ensemble method using Grid search.
2. Choose the best method through model evaluation.
3. Learn selected ensemble classifier using a small amount of labeled dataset
4. Classify m unlabeled data using learned classifier $f_{Ensemble}(X^U)$
   $X^U$ = unlabeled data
5. Combine the classification result with unlabeled data and use it as labeled data
   $f_{Ensemble}^{new} = (X^L + X^U, \quad Y^L + f_{Ensemble}(X^U))$
6. Repeat 3-4 until the parameter converges

Fig. 10. EM-based semi-supervised learning process with Ensemble learning

*E. The ranking stage*

In this section, we explain how to rank the classification results of the Ensemble model. The methods of assigning the ranking to data includes a method of calculating the distance among data [7, 8, 9] and a method of using the conditional probability [10]. In this paper, the number of common features of the aggregated open data and the support data is too small to use the distance-based ranking method. Therefore, in this paper, we adopt the conditional probability-based ranking method. Equation (2) shows the equation for conditional independent probability.

$$\Pr(y|x) = \frac{Pr(y) \prod_{i=1}^{n} Pr(x_i|y)}{\prod_{i=1}^{n} Pr(x_i)} \quad (2)$$

In this equation, $\Pr(y|x)$ represents the probability that $y$ occurs when given the features of $x$. In the previous section, $x$ can be features of support building data, and $y$ can be a feature of aggregated fire open data. We can select the most probable data among the subset of support data by using the conditional independent probability.

V. EMPIRICAL RESULTS

*A. Data description*

To evaluate the performance of the proposed method, we collected aggregated open data related to the recent Korean fire accidents (NFDS) and the support data related to building information. Table 1 shows the information of the aggregated open data used in our experiment.

TABLE I. SUMMARY OF OPEN DATA SOURCES

| Name | Number of Features | Number of instances | Description |
|---|---|---|---|
| NFDS Fire incidents dataset | 15 | 19,329 | Fire incidents from 2007 - 2017 |
| Building dataset | 27 | 634,757 | Information about buildings |
| Electricity consumption data | 2 | 634,757 | Electricity consumption per building 2016 |
| Gas consumption data | 2 | 634,757 | Gas consumption per building 2016 |

In Table 1, the 'NFDS Fire incidents' dataset is an aggregated open data based on the Fire Safety Center from January 1, 2007 to January 31, 2017, which includes fire and building related features (date, fire damage amount, building structure, purpose of building, and etc.). Next, the 'Building' dataset is the support data that includes building-related features including address, building purpose, building structure, roof structure, number of floors, and the date of construction permission. The 'Electricity consumption' data and 'Gas consumption' data are aggregated open data on electricity usage and gas consumption per building for 2016.

*B. Experiment and evaluation*

In this paper, to estimate the original building data where the fire occurred, we first found similar data in NFDS Fire incidents dataset and Building dataset, and then we performed LSH to construct it as a subset of support data; the LSH is performed while changing the number of bands and rows as described in Section 4, where the bands are {4, 5, 10, 20} and the rows are {25, 20, 10, 5}. Figure 11 shows the LSH procedure. In this figure, each data in the NFDS dataset is processed by LSH for the entire data in the 'Building' dataset, and as a result, one fire event recorded in the NFDS dataset can be associated with one or several data in the 'Building' dataset.

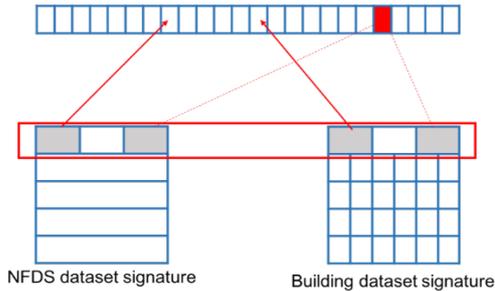

Fig. 11. LSH process between NFDS dataset and Building dataset

Figure 12 shows the probability that two data will be assigned to the same bucket according to the parameters set in this paper. We set the number of bands and rows so that two data with the similarity of 0.9 or higher can be assigned to the same bucket with high probability. Then, if there is no data in the same bucket, we perform the LSH by adjusting the number of bands and rows so that the data with relatively lower similarity can be mapped to the same bucket with high probability.

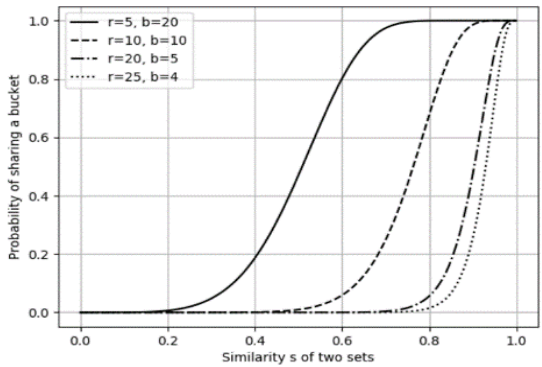

Fig. 12. Changes of the LSH performance by the number of bands and rows

Table 2 shows part of the LSH results between the 'NFDS Fire' dataset and the 'Building' dataset. The column 'NFDS' refers to data recorded in the NFDS dataset by fire event, and the column 'candidate_buildings' denotes the identifiers of the building data similar to the data in the column 'NFDS'. The column 'score' means the value of Jaccard similarity between two data. According to Table 2, there can be one or more building data similar to a one fire event.

TABLE II. SAMPLE OF THE LSH RESULTS

| NFDS | candidate_buildings | score |
|---|---|---|
| Fire_incident_1 | building_1007591 | 0.909091 |
| Fire_incident_1 | building_1007592 | 0.909091 |
| Fire_incident_2 | building_1006850 | 0.759253 |
| Fire_incident_3 | building_101441 | 0.909091 |
| Fire_incident_3 | building_101447 | 0.909091 |

In our experiments, we have performed the preprocessing of datasets and the generation of training data to obtain more correct classification model; for instance, we normalized highly skewed features among the continuous features of datasets and removed redundant features from the features of datasets. Also, we treated missing data as mean values within the same category.

The data used for model learning contains 1,653 building data where the fire occurred and 19,040 building data where the fire not occurred. We collected the data about the buildings where the fire occurred or did not occur (see Table 2) even once in the result of LSH. Because, the characteristics of the LSH can guarantee that the data which do not enter the candidate buildings are stochastically 'the data about the buildings where the fire did not occur. Then, we performed over-sampling and under-sampling on the imbalanced dataset and then performed feature selection using the random forest algorithm. Table 3 shows the top-10 importance features calculated through the random forest.

TABLE III. TOP 10 FEATURES THROUGH THE RANDOM FOREST

| Rank | Top 10 features |
|---|---|
| 1 | number of buildings |
| 2 | number of ground floors |
| 3 | number_of_households |
| 4 | permission_date |
| 5 | floor_purpose |
| 6 | floor_area_ratio |
| 7 | electricity consumption |
| 8 | ground_area |
| 9 | roof |
| 10 | gas_consumption |

We evaluated the classification models after performing the three types of Ensemble learning using the grid search (see Section 4). Table 4 shows the evaluation results for the three Ensemble techniques, which shows that the stacking model outperforms the other two techniques. Thus, we combine the EM algorithm and the stacked Ensemble method for semi-supervised learning.

TABLE IV. COMPARISON OF THE TEST PERFORMANCE I

| model | accuracy | precision | recall | F1-measure |
|---|---|---|---|---|
| bagging | 0.9946 | 0.9712 | 0.8771 | 0.9217 |
| boosting | 0.9941 | 0.9359 | 0.8983 | 0.9167 |
| **stacking** | **0.9944** | **0.9918** | **0.8628** | **0.9228** |

Table 5 shows the hyper parameters of the stacked Ensemble models by grid search. The column 'alpha' denotes the elastic-net mixing parameter of the generalized linear model (GLM), and the column 'ntrees' denotes the number of trees in the random forest and gradient boosted model (GBM). The column 'max_depth' means the maximum depth of trees of random forest and GBM, and the column 'nbins' means the number of divisions when dividing numeric data into histograms. The columns 'sample_rate' and 'col_sample_rate_per' denotes the ratio of the sample and the feature when creating the tree.

TABLE V. HYPER PARAMETERS OF THE STACKED ENSEMBLE MODEL

| Level | model | alpha | ntrees | max_depth | nbins | sample_rate | col sample rate per tree |
|---|---|---|---|---|---|---|---|
| Base model | Random Forest | - | 110 | 14 | 512 | 0.76 | 0.83 |
| | Random Forest | - | 100 | 18 | 16 | 0.86 | 0.38 |
| | GBM | - | 100 | 20 | 256 | 0.76 | 0.81 |
| | GBM | - | 110 | 20 | 512 | 0.64 | 0.65 |
| | GBM | - | 110 | 21 | 512 | 0.46 | 0.88 |
| | GLM | 0 | - | - | - | - | - |
| | GLM | 0.5 | - | - | - | - | - |
| Meta learner | GLM | 0.5 | - | - | - | - | - |

Table 6 shows the performance comparison between learning by using only stacked Ensemble and learning by combining stacked Ensemble with EM algorithm. Table 6 shows that the stacked Ensemble combined with EM method has a slightly lower precision than the only stacked ensemble method, whereas the recall and F1-score values are improved by 0.028 and 0.007, respectively. Therefore, the stacked ensemble combined with EM method outperforms the supervised learning method where only stacked Ensemble method using only the labeled dataset.

TABLE VI. COMPARISON OF THE TEST PERFORMANCE II

| Methods | accuracy | precision | recall | F1 score |
|---|---|---|---|---|
| stacked Ensemble | 0.994 | 0.992 | 0.863 | 0.923 |
| stacked Ensemble + EM | 0.996 (0.002) | 0.972 (-0.020) | 0.891 (0.028) | 0.930 (0.007) |

TABLE VII. MICRO DATA ESTIMATION PERFORMANCE

| | test data | Top1 | Top2 | Top3 | Accuracy |
|---|---|---|---|---|---|
| 1 | 192 | 103 | 11 | 1 | 0.5989 |
| 2 | 186 | 93 | 16 | 3 | 0.6022 |
| 3 | 188 | 91 | 15 | 5 | 0.5904 |
| 4 | 186 | 95 | 15 | 2 | 0.6021 |
| 5 | 185 | 94 | 15 | 2 | 0.6 |
| 6 | 186 | 84 | 12 | 11 | 0.5752 |
| 7 | 185 | 86 | 18 | 4 | 0.5837 |
| 8 | 187 | 89 | 17 | 5 | 0.5936 |
| 9 | 189 | 92 | 14 | 7 | 0.5979 |
| 10 | 186 | 88 | 15 | 8 | 0.5968 |

Next, we assigned the ranking to each data by using conditional probability in the classification results, and selected Top-3 among them. In our experiment using the conditional probability (see Equation 2), $y$ is the amount of fire damages, and $x$ is the features of the aggregated open data about buildings.

Table 7 shows the micro data estimation performance when the proposed method was conducted 10 times. In this table, the size of the test data is different for each run because the seed value of the classification model is changed by performing new EM stage and ranking stage. Table 7 show that the average accuracy when performing 10 runs is 0.5941; that is, six of the buildings that corresponded to 10 fire cases were correctly identified by the proposed micro data estimation technique.

## VI. CONCLUSION

In this paper, we proposed a method of estimating individual micro data from aggregated open data based on semi-supervised learning and conditional probability. The proposed method estimates the data that approximates the individual micro data from the aggregated open data using LSH, and combines EM algorithm with the Ensemble model to identify the data approximate to the individual micro data. Then, to select top-N from classification results, the proposed method calculates the ranking of data by using conditional probability. According to the experimental results, the proposed method shows the performance of estimating the individual micro data at about 60% in terms of accuracy. Thus, the proposed method shows that the individual micro data can be estimated at a reasonably high level from the aggregated open data. Hence, we expect that the proposed method can greatly enhances the utility of open data.

ACKNOWLEDGMENT

This research was supported by Basic Science Research Program through the National Research Foundation of Korea [grant number NRF-2015R1D1A1A09061299].